\documentclass[sigconf]{acmart}
\renewcommand\footnotetextcopyrightpermission[1]{} 
\usepackage{amsthm}
\newtheoremstyle{mystyle}
  {}
  {}
  {\itshape}
  {}
  {}
  {.}
  { }
  {\textbf{\thmname{#1}\thmnumber{ #2.}}\thmnote{ #3}}
\theoremstyle{mystyle}

\newtheorem{Definition}{\upshape{Definition}}

\usepackage{subcaption}
\usepackage{multirow}
\usepackage{multicol}
\usepackage{makecell}
\usepackage{algorithm}
\usepackage{algorithmic}

\begin{document}

\title{Heterogeneous Information Network based Default Analysis on Banking Micro and Small Enterprise Users}

\author{Zheng Zhang\authornote{Equal contributions.}\authornote{Corresponding authors.}\footnotemark[1], 
Yingsheng Ji\footnotemark[1]\footnotemark[2], 
Jiachen Shen, 
Xi Zhang\footnotemark[2]
and Guangwen Yang \\}

\email{
{zhang.zh0707, jiyingsheng, js2111jiachenshen}@gmail.com}
\email{
zhangx@bupt.edu.cn,
ygw@tsinghua.edu.cn}

\renewcommand{\shortauthors}{Zheng Zhang,Yingsheng Ji, Jiachen Shen, JIng Hong, Xi Zhang, Guangwen Yang}
\renewcommand{\authors}{Zheng Zhang,Yingsheng Ji, Jiachen Shen,JIng Hong, Xi Zhang, Guangwen Yang}

\begin{abstract}
  Risk assessment is a substantial problem for financial institutions that has been extensively studied both for its methodological richness and its various practical applications. With the expansion of inclusive finance, recent attentions are paid to micro and small-sized enterprises (MSEs). Compared with large companies, MSEs present a higher exposure rate to default owing to their insecure financial stability. Conventional efforts learn classifiers from historical data with elaborate feature engineering. However, the main obstacle for MSEs involves severe deficiency in credit-related information, which may degrade the performance of prediction. Besides, financial activities have diverse explicit and implicit relations, which have not been fully exploited for risk judgement in commercial banks. In particular, the observations on real data show that various relationships between company users have additional power in financial risk analysis. In this paper, we consider a graph of banking data, and propose a novel HIDAM model for the purpose. Specifically, we attempt to incorporate heterogeneous information network with rich attributes on multi-typed nodes and links for modeling the scenario of business banking service. To enhance feature representation of MSEs, we extract interactive information through meta-paths and fully exploit path information. Furthermore, we devise a hierarchical attention mechanism respectively to learn the importance of contents inside each meta-path and the importance of different metapahs. Experimental results verify that HIDAM outperforms state-of-the-art competitors on real-world banking data. 
\end{abstract}



\maketitle

\section{Introduction}

Micro and Small Enterprises, called MSEs for short, serve as an indispensable component of the modern economic society. In China, they not only contribute 60\% of the GDP and 50\% of the tax revenue but maintain 80\% of the total employment as well. Inspite of the backbone in economy, MSEs present a higher default risk than large companies. This is because most MSEs are fragile to external impacts and their financial statuses are uncertain and unstable. Under the auspices of national "common prosperity" objectives, banks strive to develop financial inclusion for MSEs, which are lack of verifiable credits, guarantees or collaterals. Massive inclusive loans are expected to put a heavy accuracy burden on risk assessment, which raises the advance of default analysis techniques.

\begin{figure*}[ht]
\begin{minipage}{\textwidth}
     \begin{subfigure}{0.48\textwidth}
         \vspace{0pt}    
         \centering
         \includegraphics[height = 5.5cm]{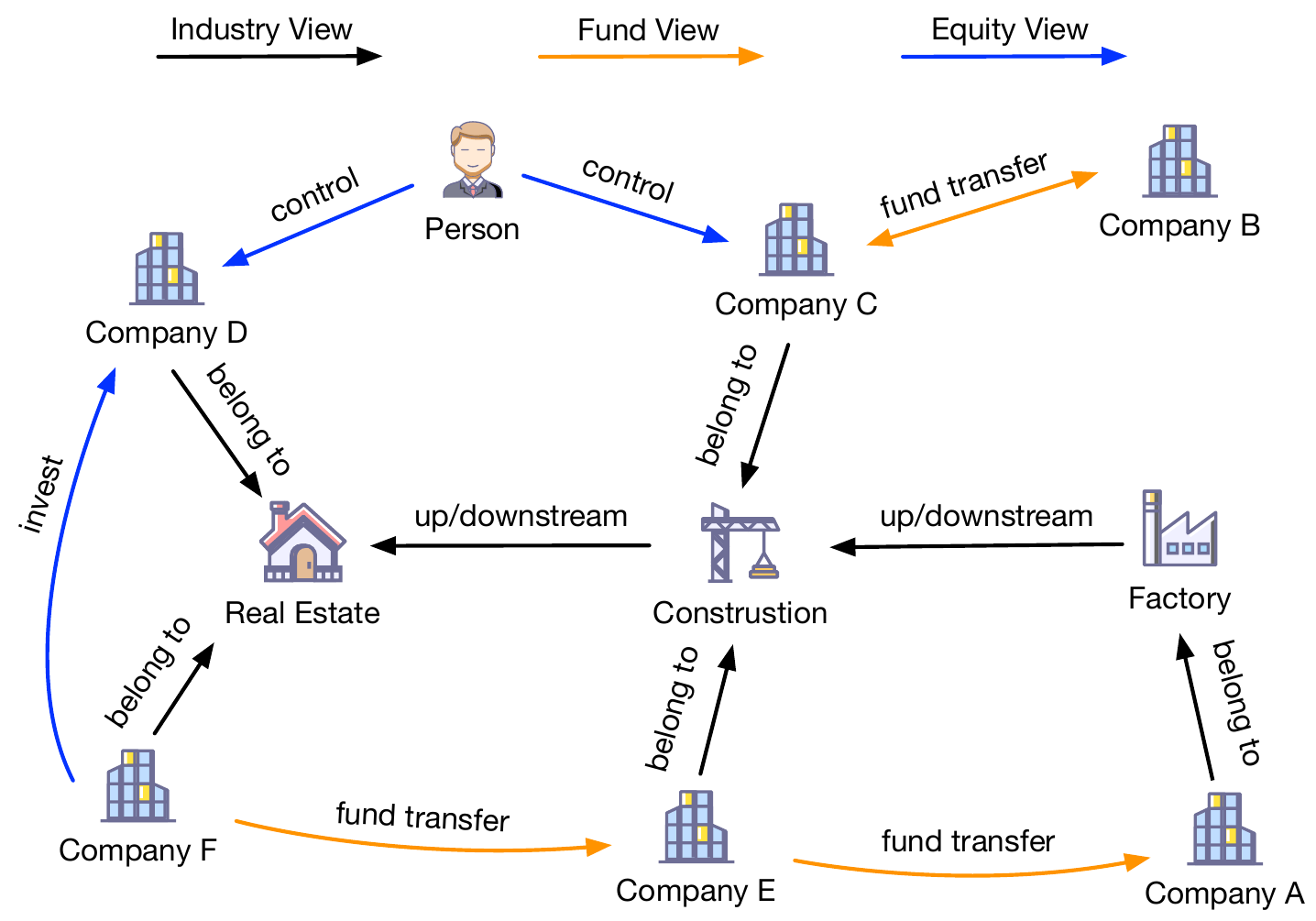}
         \caption{Banking scenario}
         \label{fig:1a}
     \end{subfigure}\hspace*{\fill}
     \begin{subfigure}{0.48\textwidth}
         \centering
         \includegraphics[height = 5.5cm]{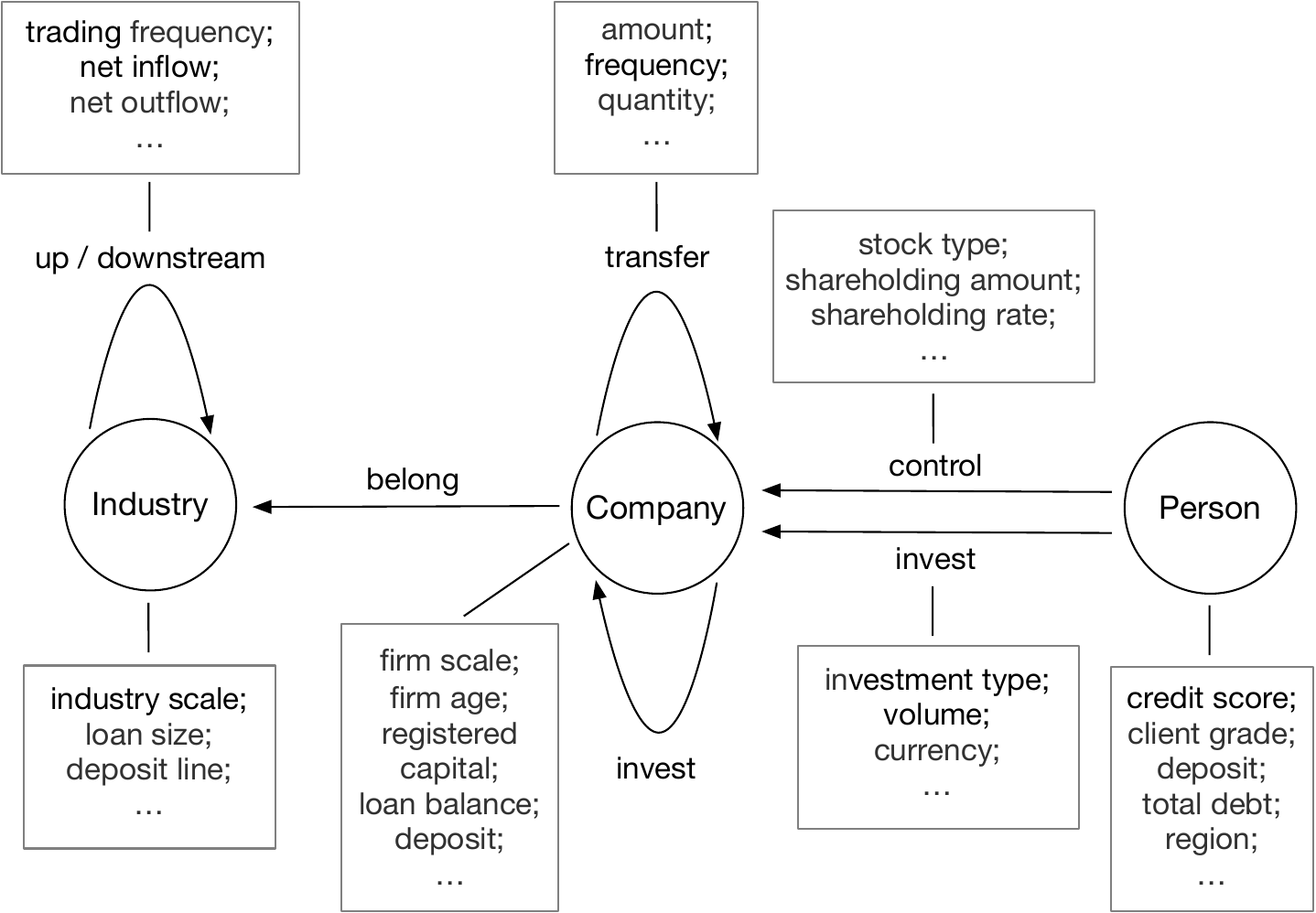}
         \caption{Network schema}
         \label{fig:1b}
     \end{subfigure}\hspace*{\fill}
\end{minipage}
\caption{The scenario of MSE default analysis.}
\label{fig1}
\end{figure*}

The intuitive approaches against potential defaults refer to rule-based classification~\cite{zhang2004discovering}. It is assumed that financial defaults have evident signals for judgement. Considering that building rules depends on prior knowledge, most rules are constantly vulnerable, i.e. they break down in face of ambiguity, variety and intricacy. To remedy such limitations, industrial solutions build 
a machine learning classifier or scorecard from historical data~\cite{lin2011machine,babaev2019rnn}. The main work is to represent a target user by extracting statistical features from diverse aspects, such as user profile, credit status, contractual capacity. Recent efforts attempt to exploit the interaction relations by using graph-based methods and verify the effectiveness for default prediction. However, most works are proposed for individual users in internet financial institutions ~\cite{yang2019understanding,wang2019heterogeneous,yang2020financial,jiang2021financial,liang2021credit}. Few studies are focused on company users in traditional commercial banks~\cite{sukharev2020ews,wang2021temporal}. The available data for graph modeling are disparate. For example, online services can provide more user action data.

Despite the remarkable successes of existing approaches and applications on various risk detection tasks, financial institutions keep developing state-of-the-art methods for more accurate and timely warning on defaults. The goal of this work is to predict whether a MSE will fail to make repayments in the future, which is not fully explored in academic and industrial communities. For this problem, we summarize the main challenges as follows:

\begin{itemize}
\item \textbf{Deficiency}: Unlike the medium or large companies, MSEs suffer from universal data deficiency. For example, most MSEs have no periodically released financial report to illustrate their business conditions, financial status and future prospect. The lack of credit-related information may significantly degrade the performance of existing models.

\item \textbf{Exogeneity}: MSE's default is not only driven by intrinsic behaviors but reflected in its associated entities as well. In a typical case, some MSE may not properly refund the debts owing to its counterparties. Traditional hand-crafted feature engineering for individual MSE cannot completely exploit a wide array of interaction relations. 

\item \textbf{Intricacy}: The default factors are comprehensive and complicated. Business is a series of social activities, which inherently presents rich sources of information for reasoning. However, massive irrelevant relations may have no contribution for prediction. It is required to capture and characterize application-dependent interactions.
\end{itemize}

To address these challenges, we propose a novel \textbf{H}eterogeneous \textbf{I}nformation network based \textbf{D}efault \textbf{A}nalysis on \textbf{M}SEs (called HIDAM for short) in this paper. Before further discussion, we take company users or companies to denote customers of the business banking service, which could be MSEs, large companies, medium companies, and else.

First of all, we model the scenario of business banking service with a heterogeneous information network (HIN)~\cite{shi2016survey, dong2020heterogeneous}. The emerging HINs offer an abstract means of representing the real-world systems generally constituted by multi-typed objects and relationships. Our intention is to leverage interaction relations and attribute information rich in financial scenarios. Inspired by the observations on real data, we integrate multiple company relationships for default analysis. Figure~\ref{fig1} exhibits an example of the well-established network in our considered scenario, which contains three views: fund view denoting money transfer relations, industry view denoting upstream and downstream relations, and equity view denoting holding relations. Considering that both objects and relationships may have respective attributes, we further incorporate an attributed heterogeneous information network (AHIN) for modeling. 

Second, we exploit node contents and composite relations on the AHIN by proposing a novel heterogeneous graph neural network. The idea is to reinforce the representation of MSEs through leveraging information involved in relations, i.e. with a set of assigned meta-paths on three views (see Table 1). The general use of meta-paths is to guide the selection of neighbors without caring information on the intermediate nodes and links. In our framework, all the messages along meta-paths are engaged in the generation of node embeddings. Specifically, we apply a linear layer for nodes and links of each type to eliminate the heterogeneity and project all node and link features to a shared feature space. Then, we elaborately devise a hierarchical attention mechanism to learn node embeddings towards heterogeneous contents. The first layer aggregates the messages inside each meta-path, while the second layer models the object preferences over meta-paths for this problem. Finally, the default probability is predicted based on aggregated information representation by a downstream multi-layer perceptron.

\renewcommand{\arraystretch}{1.2}
\begin{table*}[ht] 
\caption{Meta-paths and semantics.}
\centering
\begin{tabular}{c|c|c|c} 
\toprule 
\textbf{View} & \textbf{ID} & \textbf{Meta-path} & \textbf{Semantics}  \\ 
\hline
\multirow{2}{*}{Fund} & \makecell[c]{$CtC$} &\makecell[l]{Company $\xrightarrow{transfer}$ Company} & \makecell[l]{with direct fund transfer}\\
\cline{2-4}
& \makecell[c]{$CtCtC$} & \makecell[l]{Company $\xrightarrow{transfer}$ Company $\xrightarrow{transfer}$ Company} &  \makecell[l]{with indirect fund transfer} \\
\hline
\multirow{3}{*}{Equity} 
& \makecell[c]{$CcPcC$} & \makecell[l]{Company $\xrightarrow{control^{-1}}$ Person $\xrightarrow{control}$ Company}  & \makecell[l]{with the same controller}\\
\cline{2-4}
& \makecell[c]{$CiC$} & \makecell[l]{Company $\xrightarrow{invest}$ Company} & \makecell[l]{with direct shareholding \\ relationship} \\
\hline
\multirow{3}{*}{Industry} 
& \makecell[c]{$CbIbC$} & \makecell[l]{Company $\xrightarrow{belong}$ Industry $\xrightarrow{belong ^{-1} }$ Company} & \makecell[l]{belonging to the same \\ industry} \\
\cline{2-4}
& \makecell[c]{$CbIuIbC$} & \makecell[l]{Company $\xrightarrow{belong}$ Industry $\xrightarrow{up/downstream}$ Industry  $\xrightarrow{belong ^{-1} }$  Company} & \makecell[l]{belonging to the \\ up/downstream industry} \\

\bottomrule 
\end{tabular}

\label{tab:meta-path} 
\end{table*}

In summary, our main contributions are as follows:

\begin{itemize}
\item To our knowledge, we are among the first attempts to address the financial default analysis on MSE users, which is an immediate issue concerned by commercial banks.
\item We quantitatively explore banking data and define an AHIN for representing rich interactions of company users and model the given problem as a node classification. 

\item We propose a novel model HIDAM to predict defaults, which is equipped with a class of meta-path based aggregations to characterize MSEs and a hierarchical attention mechanism to adaptively capture the key information involved in the local structure and meta-paths.
\item We conduct experiments on real-world data, hosted by a top financial institution. The results verify that HIDAM outperforms other competitors and has good interpretability. 
\end{itemize}

Furthermore, HIDAM provides natural extensibility and can easily incorporate more relations and features for other banking risk scenarios. Our model is reproducible, and open-sourced\footnote{https://github.com/adlington/HIDAM}.

\section{Related Work}

In this section, we summarize previous works from two aspects: data and methods. Note that financial risk assessment, related to user behavior, can be associated with credit scoring of new loans, default prediction of existing loans, and fraud detection owing to higher exposure to default. 

\subsection{Graph Data for Risk Analysis}

Recent efforts attempt to incorporate diverse interactions for graph-based credit risk analysis. Transactional data serve as the footstone of building financial networks.~\citet{sukharev2020ews} focused on the social connections from purchases and money transfers to improve the quality of credit scoring. Social relationships are rich in transfer transaction scenarios, such as relatives, friends, colleagues between node pairs~\cite{wang2021temporal}. Rule-based aggregated transactions can provide more structural and semantic information, e.g. shared-card network between merchants for fraud detection~\cite{khazane2019deeptrax}. Besides, a few studies exploited social behaviors from user submitted data to identify micro-credit frauds, such as mobile network from call logs~\cite{yang2019understanding}, residential location embedding network~\cite{yang2019understanding}. Guarantee network can provide powerful features for risk prediction~\cite{cheng2019risk}. This is because guarantee for easy-loans will increase the coupled dependencies and may magnify single default risk to contagious chain risk. Most closely related to ours is the supply-chain graph for risk analysis~\cite{yang2020financial}, which employs a semi-supervised link prediction to mine supply chain relationship and a supervised node classification to estimate loan default. The supply relations are calculated on real money transfer while our adopted industrial chain is focused on the logistic coordination among industries and companies. To our knowledge, this is the first work of graph-based heterogeneous data mining for credit risk analysis in commercial banks.

\subsection{Heterogeneous Information Network}

As a powerful modeling method, HIN can explicitly distinguish multiple object types and relationship types for representing real systems. Compared to the broadly-used homogeneous information network, this method can integrate more complex structures and more abundant semantics for downstream graph tasks. HINs have been an important focus of graph communities and has been widely applied in many domains: recommender system~\cite{hu2018leveraging, fan2019metapath, zhao2019motif, liu2021heterogeneous}, malware mitigation~\cite{ye2019out, hou2019alphacyber}, fraud detection~\cite{liu2018heterogeneous, dou2020enhancing}, intelligent transportation~\cite{zhu2019tdp, hong2020heteta} and security~\cite{wang2019spotting, ji2021prohibited}. In financial scenarios,~\citet{hu2019cash} considered node information and proposed an attributed heterogeneous information network (AHIN) for cash-out fraud detection.~\citet{hu2020loan} exploited multiplex relations for spotting loan defaults, as is a specific heterogeneous network.~\citet{zhong2020financial} focused on default users by exploring multi-view AHIN.~\citet{liang2021credit} considered credit risk and limits forecasting as a multi-task learning framework through multiple views of heterogeneous information. However, most risk assessments are designed for personal finance and interest financial scenarios, such as mobile credit payment, internet loan, E-Commerce consumer lending service. Few works refer to default analysis on banking micro and small company users, the risk characteristics of which have not been extensively studied.

\section{Preliminaries}

In this section, we first present the problem statement of our work. After that, we explore data to support our intuitions.

\subsection{Problem Statement}

We represent rich information derived from the scenario of business banking service as an attributed heterogeneous information network. The formal definition is as follows. 

\begin{Definition} [\textbf{Banking Company Netwrok} (BCN)]
\emph{A BCN is denoted as $G=\{V, E, X_{V}, X_{E}\}$, consisted of nodes $V$ and links $E$ with attributes $X_{V}=\left\{X_{A_{i}}\right\}$ and $X_{E}=\left\{X_{R_{i}}\right\}$, respectively. Here, $X_{A_{i}} \in \mathbb{R}^{\left|V_{A_{i}}\right| \times d_{A_{i}}}$ is a node attribute matrix with $\left|V_{A_{i}}\right|$ nodes and $d_{A_{i}}$ attributes. Correspondingly, $X_{R_{i}} \in \mathbb{R}^{\left|E_{R_{i}}\right| \times d_{R_{i}}}$ is a link attribute matrix with $\left|E_{R_{i}}\right|$ links and $d_{R_{i}}$ attributes. Let $\mathcal{A}=\left\{A_{i}\right\}$ and $\mathcal{R}=\left\{R_{i}\right\}$ denote the sets of predefined node and link types, where $\left|\mathcal{A}\right|+\left|\mathcal{R}\right|>2$.} 
\end{Definition}

As shown in Figure~\ref{fig1}, we give an illustrative example of the BCN containing three main types of nodes (i.e. Company (C), Person (P), Industry (I)) with respective attributes and relations. 

We adopt meta-paths to capture the complex structural and semantic information in BCN. A meta-path is formed as an ordered sequence composed of multi-typed nodes and links, which is aimed to express composite relations between two objects. Formally, 

\begin{Definition}[\textbf{Meta-path}]
\emph{A meta-path $\mathcal{P}$ is defined in the form of $A_{1} \stackrel{R_{1}}{\rightarrow} A_{2} \stackrel{R_{2}}{\rightarrow} \cdots \stackrel{R_{l}}{\rightarrow} A_{l+1}$, which depicts a collection of relations $R_{1} \circ R_{2} \circ \cdots \circ R_{l}$ between $A_{1}$ and $A_{l+1}$, where $\circ$ is the composition operator on relations.}
\end{Definition}

Take Figure~\ref{fig:1a} as an example, two companies can be connected via diverse semantic paths: the $CtC$ path indicates that two companies have transactions, the $CcPcC$ path  indicates that two companies have the same controller, and the $CbIuIbC$ path indicates that two companies are in upstream/downstream relation industries, as listed in Table~\ref{tab:meta-path}. For simplicity, a concrete instance of meta-path is called \emph{path instance}, which connects the given node with a specific node called \emph{meta-path based neighbor}.

Based on the proposed BCN, we further formalize our financial default analysis problem as follows. 

\begin{Definition}[\textbf{Default analysis under BCN}]
\emph{Given a target set of MSEs, denoted as $U$, our purpose is to infer the binary label $y_{u} \in\{0,1\}$ for each MSE $u\in U$ to indicate whether $u$ will be default over a period of time. Note that the default analysis task is based on attribute and interaction information distilled from the BCN $G=\{V, E, X_{V}, X_{E}\}$ and hence $U$ is a subset of nodes (i.e. $U \subseteq V$).}
\end{Definition}

Here, we concentrate on the MSE users that recently obtain their first loans from the bank, which suffer obvious credit-related data deficiency problem.

\begin{table}[t]
\caption{Statistics of banking relational data.}
\centering
\begin{tabular}{l|l|r} 
\toprule
\textbf{View}   &   \textbf{Relation}   &   \textbf{Coverage}   \\
\hline
\multirow{1}{*}{Fund}     & (Company, transfer, Company)    & 50.3\%    \\
\hline
\multirow{3}{*}{Equity}   & (Company, invest, Company)      & 25.2\%    \\
                          & (Person, control, Company)      & 90.4\%    \\
                          & (Person, invest, Company)       & 5.2\%     \\
\hline
\multirow{1}{*}{Industry} & (Company, belong, Industry)     & 95.1\%    \\
\hline
\multirow{2}{*}{Business} & (Company, guarantee, Company)   & 0.6\%     \\
                          & (Company, hasbill, Company)     & 4.5\%     \\

\bottomrule
\end{tabular}
\label{tab:coverage} 
\end{table}

\begin{table}[t] 
\caption{Data deficiency of MSEs and the effects of information supplement from view-specific neighbors.}
\centering
\begin{tabular}{l|r|r|r|r|r} 
\toprule 
 \multirow{2}{*}{\textbf{Feature set}}
 & \multirow{2}{*}{\textbf{MSE}}
 & \multicolumn{4}{c}{\textbf{MSE filled by neighbors}} \\ 
 \cline{3-6}
 & & 
 \multicolumn{1}{r|}{\textbf{Equity}} & \multicolumn{1}{r|}{\textbf{Industry}} 
 & \multicolumn{1}{r|}{\textbf{Fund}} & \multicolumn{1}{r}{\textbf{All}}  \\
\hline
\textbf{User profile} & 18.4\% & 17.3\% & 15.1\% & 7.1\% & 7.0\% \\
\hline
\textbf{Credit status} & 48.7\% & 46.6\% & 41.0\% & 25.2\% & 25.0\% \\
\hline
\textbf{Solvency} & 89.5\% & 88.9\% & 83.8\% & 58.1\% & 58.0\% \\
\hline
\textbf{Operation} & 55.3\% & 54.4\% & 48.9\% & 22.4\% & 22.3\% \\
\hline
\textbf{Activity} & 7.1\% & 6.0\% & 4.7\% & 1.6\% & 1.4\% \\
\bottomrule 
\end{tabular}
\label{tab:view-deficiency} 
\end{table}

\subsection{Exploratory Analyses}

We conduct data investigation into the following perspectives of the graph-based default prediction problem: \textbf{(P1)} What kinds of relational data are used for graph modeling? \textbf{(P2)} What is the status of financial data closely related to risk assessment? \textbf{(P3)} What is the correlation between relationships and default risks?

\textbf{Connection analysis (P1).} We first explore a variety of financial relationships in the business banking service. Our purpose is to find the relational data used for comprehensively connecting company users. The exploration is performed on 17.1 millions of companies and 182.3 millions of relations. For each specific relationship, we count the number of involved companies over the total number of companies, which is called coverage. As shown in Table~\ref{tab:coverage}, we can observe that the relationships from business view are obviously sparse, which may incur a serious underfitting problem. By comparison, the vast majority of companies are connected through three views, i.e. fund, equity, and industry. Based on the above observations, we jointly integrate these view-specific relationships, the schema of which is detailed in Figure~\ref{fig:1b}.

\textbf{Feature analysis (P2).}  We investigate the status of banking data for expressing MSEs. The investigation is performed on five sets of features, including user profile (i.e. client and account information), credit status (i.e. credit report and blacklist), solvency (i.e. contractual capability), operation (i.e. financial performance) and activity (i.e. account behavior and transaction action). For each set, we count the number of missing values over the total attributes for each MSE and then compute the average missing rate. As shown in Table~\ref{tab:view-deficiency}, we can observe that MSEs usually suffer significant deficiency in various aspects of attribute information. 

Next, we verify that the data deficiency problem for MSEs can be mitigated through aggregating their neighbors. For this purpose, we count missing attributes of a target MSE together with the meta-path based neighbors. To reveal the effects of different neighbors, five experimental groups are defined for comparison, including MSEs themselves, MSEs with neighbors guided by meta-paths from a single view (equity, industry, fund), and MSEs with neighbors from three views. As shown in Table~\ref{tab:view-deficiency}, the results exhibit that feature sets obtain enhancement to varying degree, where the average missing rates decrease by 80\% at most. We can observe that the relationships, via meth-paths from each view, improve the information of MSEs and all the relationships achieve the best average missing rates. 

\begin{figure}[t]
    \begin{subfigure}{0.23\textwidth}
         \centering
         \includegraphics[scale = 0.2]{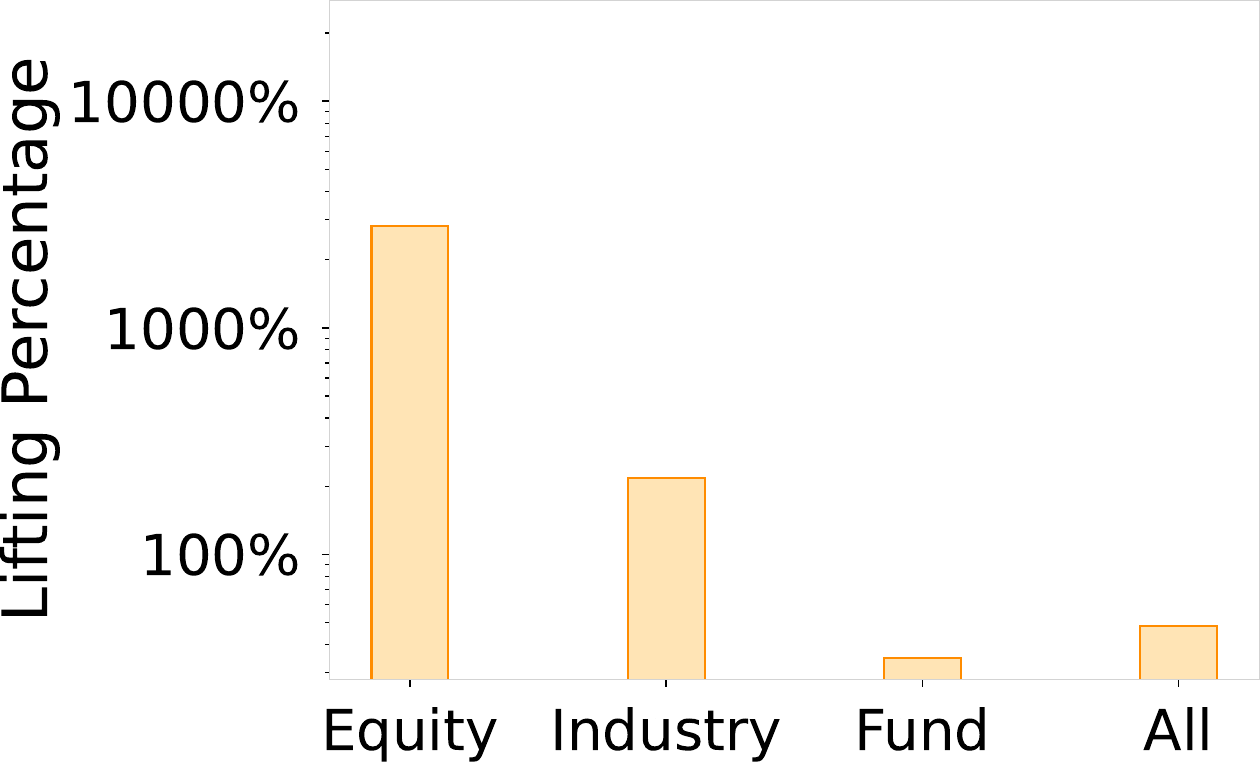}
         \caption{Lifting percentages of default}
         \label{fig:3a}
     \end{subfigure}\hspace*{\fill}
     \begin{subfigure}{0.23\textwidth}
         \centering
         \includegraphics[scale = 0.2]{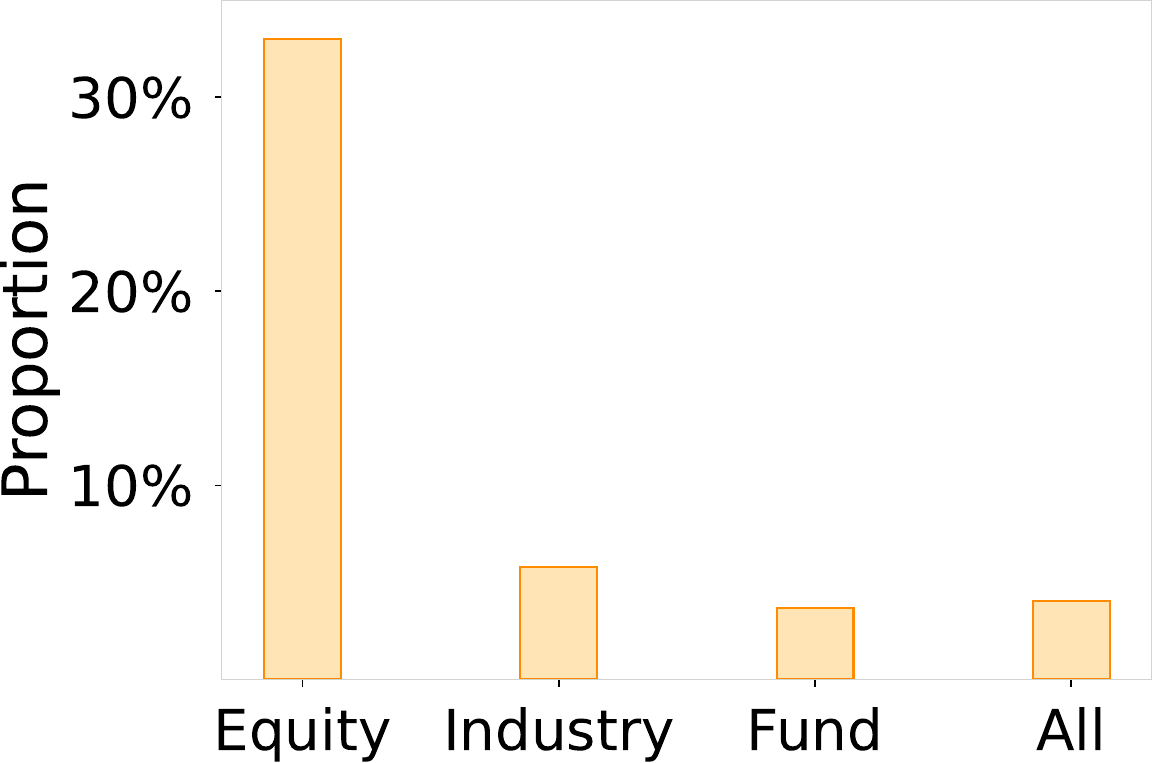}
         \caption{MSEs with default neighbors}
         \label{fig:3b}
     \end{subfigure}\hspace*{\fill}
\caption{Analysis on MSEs with neighbors from different views. (a) shows the lifting percentages of default proportion in MSEs with default neighbors against MSEs with no default neighbors. (b) shows the proportion of default MSEs with default neighbors in MSEs with neighbors.}
\label{fig2}
\end{figure}

\begin{figure*}[t]
\begin{minipage}{\textwidth}
     \centering
     \includegraphics[scale = 0.22]{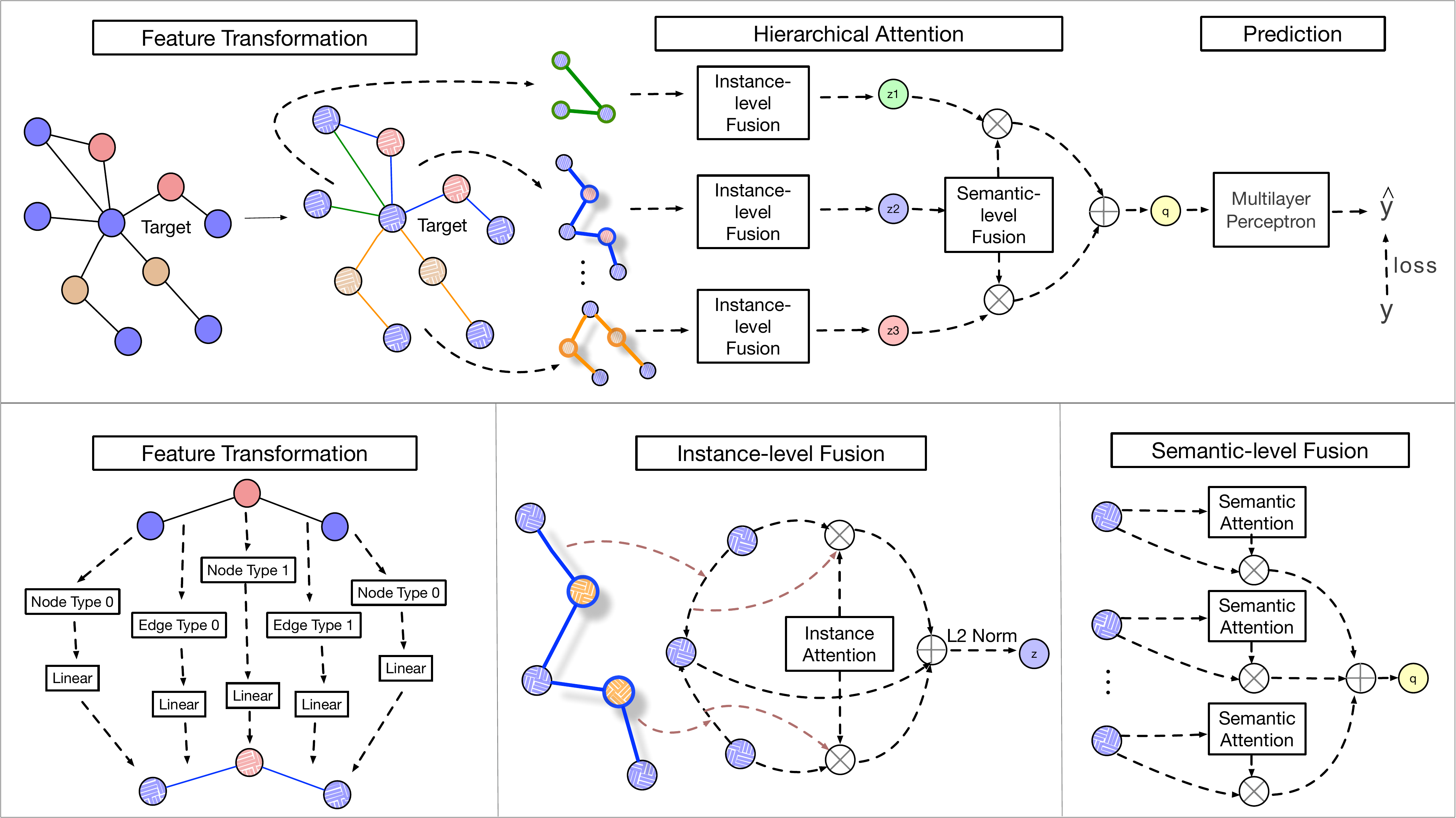}
\end{minipage}
\centering
\caption{The architecture of the proposed model.}
\label{fig4}
\end{figure*}

\textbf{Impact analysis (P3).} We further study the correlation between relationships and default risks. For each view, we collect the meta-path based neighbors of each MSE and count the number of default neighbors. All the MSEs are divided into two groups: MSEs with default neighbors and MSEs with no default neighbors. We count the proportion of default MSEs in each group, and then calculate the lifting percentages of default proportion in MSEs with default neighbors against MSEs with no default neighbors. The comparative results between two groups are demonstrated in Figure~\ref{fig:3a}. We can find that three views with relationships have distinct impacts on MSEs' default risks. Furthermore, we explore the connectivity between default MSEs. For each view, we count the number of default MSEs with default neighbors over the total number of default MSEs having neighbors. In Figure~\ref{fig:3b}, we can find that some default MSEs have consistent risks with their neighbors, especially equity view, where MSEs are obviously influenced by default neighbors. The findings imply that the relationships (i.e. meta-paths) from three views are likely effective for default prediction of MSEs. 

\section{The Proposed Model}

In this section, we study heterogeneous graph learning and propose the model of HIDAM. First, we integrate the structural characteristics for each MSE through a set of meta-paths. To address the heterogeneity, we convert the node and link attributes into a unified-size feature space via feature transformation. Then, we design a hierarchical attention mechanism to characterize the risk profile of MSEs. Instance-level fusion learns features among different path instances of each assigned meta-path consisted of multi-type attributed nodes and links. Semantic-level fusion captures semantic importance from associated meta-paths. Finally, HIDAM combines node embeddings with the downstream learning objective for prediction. Figure~\ref{fig4} shows the overall architecture of our model. 

\subsection{Meta-paths in Business Scenarios} 

We adopt meta-path to capture the structures of BCN, i.e. high-order proximity between two nodes. Meta-path can effectively capture the complex semantics or relations among objects and has been well used for various risk detection tasks~\cite{cao2017hitfraud,hou2019alphacyber,hu2019cash,zhong2020financial}. Recall that we have verified that the company relationships are conductive to tackling the default prediction. In particular, the data deficiency for MSEs is obviously decreased by absorbing their meta-path based neighbor information. The meta-paths from different views have diverse influence on the default risks of MSEs.

Motivated by the above observations, we aggregate nodes and links for each MSE through a set of elaborately devised meta-paths. Table~\ref{tab:meta-path} exhibits the meta-paths and their semantics. Note that we customize the meta-paths based on available data and expert knowledge. Our proposed meta-paths are closely related to the default risk of MSEs. For example, the $CbIuIbC$ path refers to the correlation between MSEs and their upstream/downstream companies. It implies that the MSE could not get the money from its suppliers, which might significantly affect its financial stability. Likewise, the counterparties with large transactions have certain influence since MSEs are fragile and their financial status is uncertain, i.e. the $CtC$ path. Furthermore, we believe that the actual controller has a huge impact on MSEs. The risk of MSEs with the same controller is also highly correlated, which derives the $CcPcC$ path.

\subsection{Feature Transformation}

The module is to project input data into the unified latent feature space. In our framework, the inputs involve a series of node and link contents assigned by meta-paths, which may have disparate feature vectors. To overcome such heterogeneity of attributes, the input layer is to integrate the original data into a uniform feature space. Concretely, we adopt to represent a category of nodes and links via a simple linear transformation. Let $d$ be the parameter to fix the size of node and link embeddings. Given a node type $A \in \mathcal{A}$ and a link type $R \in \mathcal{R}$, for each node $v \in V_{A}$ and each link $e \in E_{R}$, we define the prejection as below:

\begin{equation}\label{eqn:1}
h_{v}=W_{A}\cdot x_{v}, h_{e}=W_{R}\cdot x_{e}
\end{equation}

\noindent where $x_{v} \in X_{A}$ and $x_{e} \in X_{R}$ are the initial vectors, $h_{v}\in \mathbb{R}^{d}$ and $h_{e} \in \mathbb{R}^{d}$ are the feature vectors of nodes and links in the hidden space, $W_{A} \in \mathbb{R}^{d \times d_{A}}$ and $W_{R} \in \mathbb{R}^{d \times d_{R}}$ are trainable weight matrices.

\subsection{Hierarchical Attention}

After transformation, nodes and links share the same latent representation. Next, we consider how to represent the path instance, the aim of which is to take full advantage of both node and link content features along the meta-path except two end nodes. Formally, given a meta-path $P$, for a node $u \in V$, we can obtain its meta-path based neighbor set $\mathcal{N}_{u}^{P}$. Let $p_{u v}$ be a specific path instance, where $v \in \mathcal{N}_{u}^{P}$. Let 
$M=\left\{m^{p_{u v}}\right\}$ be the intermediate nodes and links of $p_{u v}$, which is assumed to have $L$ elements, i.e. $\left|M\right|=L$. We further encode the path instance through a fusion function $f(\cdot)$: 

\begin{equation}\label{eqn:2}
h_{p_{u v}}=f\left(\left\{h_{m} \mid \forall m \in\left\{m^{p_{u v}}\right\}\right\}\right)
\end{equation}

\noindent Here, we adopt a simple approach to directly concatenate the intermediate feature vectors. The output $h_{p_{u v}}$ can be viewed as a vector containing features of a virtual link between $u$ and $v$. 

So far, we have obtained the transformed node contents and encoded their meta-path based messages into feature vectors. Next, we exploit these heterogeneous contents for each target node in two hidden layers. In brief, we design a hierarchical attention structure: instance-level fusion and semantic-level fusion to model the preferences over the local information as well as metapaths. The distillation procedure can be visualized in Figure~\ref{fig4}.

\textbf{Instance-level fusion}. Intuitively, multi-type nodes and links of a path are likely to have different importance. For example, the neighbors with large transactions may have larger impacts on the MSE's financial status. Hence, we integrate the node and link feature contents within each meta-path separately. Inspired by~\cite{brody2022how} and real data observations, we adopt a dynamic attention layer to weight and aggregate the path instances of $P$ for each target node $u$. The following equations represent an attention operation: 

\begin{equation}\label{eqn:3}
\alpha_{p_{u v}}=\frac{\exp \left(a_{P} \cdot \sigma\left(\left[h_{u}\left\|h_{v}\right\| h_{p_{u v}}\right]\right)\right)}{\sum_{v^{\prime} \in \mathcal{N}_{u}^{P}} \exp \left(a_{P} \cdot \sigma\left(\left[h_{u} \| h_{v^{\prime}}|| h_{p_{u v^{\prime}}}\right]\right)\right)}
\end{equation}

\begin{equation}\label{eqn:4}
\varphi_{u}^{P}=\sigma\left(h_{u}+W_{\phi} \sum_{v \in \mathcal{N}_{u}^{P}} \alpha_{p_{u v}}\left[h_{v} \| h_{p_{u v}}\right]\right)
\end{equation}

\begin{equation}\label{eqn:5}
z_{u}^{P}=\frac{\varphi_{u}^{P}}{\left\|\varphi_{u}^{P}\right\|_{2}}
\end{equation}

\noindent where $a_{P} \in \mathbb{R}^{(L+2)d}$ and $W_{\phi} \in \mathbb{R}^{(L+1) d \times d}$ are the parameterized attention vector and weight matrix for learning. For a path instance $p_{u v}$ between node $u$ and its neighbor $v$, the attention value $\alpha_{p_{u v}}$ is computed by a scoring function of the concatenated vector $\left[h_{u} \| h_{v} \| h_{p_{u v}}\right]$ and then normalized by a softmax unit, as is formed in Eq.~\ref{eqn:3}, where $\sigma(\cdot)$ is an activation function. The representation $\varphi_{u}^{P}$ is computed by a linear combination of transformed features within all the meta-path instances and the pre-activation residual connection is added to keep its own features~\cite{li2020deepergcn} as is formed in Eq.~\ref{eqn:4}. The final output $z_{u}^{P}$ is transformed through the L2 normalization operation to keep features at the same scale~\citep{wu2021r}.

Note that the BCN for modeling the scenario of business banking service involves various link attributes. Most of them are extracted from credit-related data and thus offer vital semantic information. This is different from recent HIN algorithms such as HAN, MAGNN. Therefore, we define the above paradigm to take advantage of both node and link attributes along the meta-path.

\textbf{Semantic-level fusion}. For a target node $u$, we have obtained the meta-path specific embeddings by the instance-level fusion. Based on the observations on real data, meta-paths may have separate contributions on the financial default. Next, we model the importance over diverse meta-paths for collaboration. Specifically, the attention weights of meta-paths are computed and aggregated by the following operations. 

\begin{equation}\label{eqn:6}
\beta_{P}=\frac{\exp \left(a_{A} \sigma\left(z_{u}^{P}\right)\right)}{\sum_{P^{\prime} \in \mathcal{P}} \exp \left(a_{A} \sigma\left(z_{u}^{P^{\prime}}\right)\right)} 
\end{equation}

\begin{equation}\label{eqn:7}
q_{u}=\sum_{P \in \mathcal{P}} \beta_{P} z_{u}^{P}
\end{equation}

\noindent where $a_{A} \in \mathbb{R}^{d}$ is the learnable meta-path preference vector and $\beta$ is the attention weights over different meta-paths. The final output $q_{u}$ is summed by a weighted average of all the meta-path specific embeddings of the target node $u$.

\subsection{Model learning}

Our proposed model is to predict the default probability that a MSE $u$ will fail to repay the loan. Recall that we obtain the distilled representation of every target MSE through aggregating the meta-path based messages and node content features. Next, we feed these fused embeddings into MLP layers as follows.

\begin{equation}\label{eqn:8}
\hat{y}_{u}=\operatorname{sigmoid}\left(w_{2} \operatorname{ReLU}\left(W_{1} q_{u}+b_{1}\right)+b_{2}\right) 
\end{equation}

\noindent Here, we adopt an element-wise rectified linear unit function, where $W_{1}$ and $b_{1}$ denote the weight matrix and the bias vector. The output prediction is obtained via a regression with a sigmoid function to ensure the output $\hat{y}_{u}$ in the range of $\left[0, 1\right]$,  where $w_{2}$ and $b_{2}$ denote the weight vector and the bias. 

We train the model with cross entropy loss with regularization. The objective function is formulated as follows:

\begin{equation}\label{eqn:9}
\mathcal{L}(\Theta)=-\sum_{\left(u, y_{u}\right) \in \mathcal{D}}\left(y_{u} \log \hat{y}_{u}+\left(1-y_{u}\right) \log \left(1-\hat{y}_{u}\right)\right)+\lambda\|\Theta\|_{2}^{2}
\end{equation}

\noindent where $y_{u}$ is the ground truth, $\Theta$ is the set of model parameters for optimization, and $\lambda$ is the regularization parameter.

\renewcommand{\arraystretch}{1.2}
\begin{table*}[t] 
\caption{The statistical information of network.}
\centering
\begin{tabular}{cccccc}
\toprule
\textbf{Dataset}  & \textbf{\#Total}  & \textbf{Type} & \textbf{Number}  & \textbf{\#Attributes} & \textbf{Example} \\
\midrule
\multirow{3}{*}{Node} & \multirow{3}{*}{6,309,269} & Company & 3,189,109 & 456 &  Age/Scale/RegisteredCapital/…  \\
 & & Person & 3,119,713 & 5 & CreditScore/CustomerType/Branch/… \\
  & & Industry & 447 & 22 & CorpNum/AvgAge/LoanSize/… \\
 \hline
\multirow{5}{*}{Link} & \multirow{5}{*}{21,842,366} & Transfer & 12,657,994 & 9 &  TradeAmount/TradeNum/TradeFreq/…  \\
 & & Belong & 5,506,600 & - & - \\
 & & Updownstream & 663 & 12 & NetInflow/NetOutflow/TradeFreq/… \\
 & & Control & 3,537,824 & 5 & RelaType/InvAmt/InvPct/… \\
 & & Invest & 139,285 & 5 & InvAmt/InvCurr/InvType/… \\
\bottomrule
\end{tabular}
\label{tab:MAHIN}
\end{table*}

\subsection{Discussion}

We further give some analysis of our proposed HIDAM. First, our intention is to build a flexible framework to leverage rich attributes and structures over the abstracted BCN from the scenario of business banking service. By elaborately devising meta-paths, we can model multiple views of data to learn comprehensive embeddings, which can enhance the default prediction for MSEs. Based on the above framework, more domain specific data and knowledge can be jointly integrated for our model. Second, our proposed model is highly efficient and applicable for large-scale networks. Specifically, HIDAM takes $O(I \cdot \sum_{P\in \mathcal{P}}(d_1 \left|V_P \right|+d_2 \left|E_P \right|+d_3 \left|R_P \right|))$ time, where $I$ is the number of epochs, $\left|V_P \right|$ denotes the number of nodes, $\left|E_P \right|$ denotes the number of edges, $\left|R_P \right|$ denotes the number of path instances, $d_1$ and $d_2$ are respectively the average attribute dimension of the nodes and links, and $d_3$ is the dimension of embeddings. Hence, the time complexity is linear to the number of nodes and edges and the number of path instances. In addition, the model can be used for inductive problems. For a new MSE which is never in training data, HIDAM can also learn the representation through its various path instances in the network.

\section{Experiments}

In this section, we conduct experiments on banking data to answer the following questions: 

\begin{itemize}
\item[\textbf{Q1.}] How accurately can our model predict real-world defaults compared to state-of-the-art competitors? 
\item[\textbf{Q2.}] How do the core components (e.g. multiple meta-paths and attention mechanism) enhance the prediction? 
\item[\textbf{Q3.}] What can be learned from the hierarchical attention mechanism?
\item[\textbf{Q4.}] Is our model sensitive to the parameters and how the performance will be affected? 
\end{itemize}

\subsection{Experimental Setup}

\textbf{Datasets.} We evaluate HIDAM on the real-world data collected from a major commercial bank in China, during a whole year. The dataset contains 158 thousand MSE users (ranging from Jan.01, 2020 to Jun.30, 2020) for training and 135 thousand MSE users (ranging from Jul.01, 2020 to Dec.31, 2020) for testing. All the MSE users obtain their first loan by bank within the given time. The positive samples are defined as MSEs that have delayed or failed in the payments of principal and interests over seven days. Totally, we collect 7 thousand default samples with the positive rate close to 2.5\% in our dataset. 

We construct the BCN based on the banking data. The whole network is consisted of 6.31 million nodes (three different types) and 21.84 million links (five different relations from three views). To enrich information for nodes, we exclude features with miss rate over 90\% and extract 456 attributes for company user node, from the aspects of user profile, credit status, solvency, operation and activity. For other types of nodes and relations, we collect a certain number of attributes ranging from 5 to 22 except belong relation. The detailed descriptions are shown in Table~\ref{tab:MAHIN}.

\textbf{Baselines.} We compare HIDAM against different kinds of methods, categorized as traditional tree-based models, homogeneous GNNs, and heterogeneous GNNs, respectively. The list of competitors is outlined as follows:  

\begin{itemize}
\item Tree-based methods: \textbf{LightGBM}~\citep{ke2017lightgbm} is a high-efficiency framework using tree-based ensemble algorithms. \textbf{DeepForest}~\citep{zhou2017deepforest} is a deep model with fewer hyper-parameters than other DNNs. We only feed node attributes of target MSEs to train a classifier. 
\item Homogeneous GNNs: \textbf{GCN}~\citep{kipf2017semi} is a deep graph convolutional network algorithm, which generates node embedding by aggregating information from neighbors.  \textbf{GAT}~\citep{velivckovic2018gat} improve the neighborhood aggregation scheme of GCN by introducing the attention mechanism. 
\item Heterogeneous GNNs: \textbf{HAN}~\citep{wang2019heterogeneous} is a graph neural network for heterogeneous graph, by incorporating node-level and semantic-level attentions for use.  \textbf{MAGNN}~\citep{fu2020magnn} employs node content transformation, intra and inter meta-path aggregations for heterogeneous graph embedding task. We feed embeddings to MLP for node classification. 
\end{itemize}

Here, we evaluate the tree-based methods with 100 trees and a maximum depth of 5, and the GNN-based methods with the number of sampled neighbors as 20, respectively. Besides, for homogeneous GNNs, we take the meta-path neighbor graph as input where all meta-paths are considered and the types are ignored.

Our implementation is based on DGL with PyTorch backend. We randomly initialize model parameters with the Xavier initializer and choose Adam as the optimizer. Moreover, we set the batch size to 512, the learning rate to 0.001 and set the weight decay to 0.01 to prevent overfitting. We also perform early stopping during the training if the validation performance is not improved for 50 epochs. In addition, the model performance is evaluated mainly via AUC (Area Under the ROC Curve) and KS (Kolmogorov Smirnov), which are extensively applied in banking scenarios. Higher scores of AUC and KS signify better default prediction for MSEs.

\renewcommand{\arraystretch}{1.2}
\begin{table*}[ht]
\caption{Performance results. The subscripts donate the increment compared to LightGBM as the performance baseline.}
\centering
\begin{tabular}{c|c|c|c|c|c|c|c}
\toprule
Metric  & LightGBM & DeepForest & GCN & GAT & HAN & MAGNN & HIDAM \\
\hline
AUC  & 0.7080$_{/0.000}$ & 0.7080$_{/0.000}$ & 0.7137$_{/0.0057}$ & 0.7177$_{/0.0097}$ & 0.7235$_{/0.0155}$ & 0.7270$_{/0.0190}$ & \textbf{0.7404}$_{/0.0324}$  \\
\hline
KS & 0.3083$_{/0.000}$  & 0.3144$_{/0.0061}$ & 0.3184$_{/0.0101}$ & 0.3253$_{/0.0170}$ &0.3398$_{/0.0315}$ & 0.3410$_{/0.0327}$ & \textbf{0.3600}$_{/0.0517}$\\
\bottomrule
\end{tabular}
\label{tab:comp}
\end{table*}

\subsection{Real-world performance (Q1)}

We compare the performance of HIDAM with the baseline methods, as is presented in Table~\ref{tab:comp}. The main observations are summarized as follows:

\begin{itemize}
\item Our model outperforms the baseline methods by a significant margin. In particular, its AUC and KS are at least 4.5\% and 14.5\% higher than the tree-based methods. Compared with the best performance of other competitors, HIDAM shows improvements with at least 1.8\% increased AUC and 5.5\% increased KS, respectively. The results verify the effectiveness of HIDAM for the default prediction, which benefits from a more principled mechanism to jointly exploit node contents and structures in the BCN.

\item Tree-based methods (i.e. LightGBM and DeepForest) exhibit relatively poor performance in predicting defaults for MSEs. As we can see, conventional feature engineering cannot take advantage of available banking data for modeling the specific and complex financial scenario, especially relational data, and therefore the performance is limited.

\item Homogeneous GNNs (i.e. GCN and GAT) obtain better performance than tree-based methods, with reported value of 0.8\% improved AUC and 1.2\% improved KS at the same time. The effects are mainly owing to the exploitation of structural information. Besides, GAT behaves better than GCN, which indicates that capturing the importance of neighbors is beneficial to capture the portrait of default MSEs.

\item Heterogeneous GNNs (i.e. HAN and MAGNN) yield competitive performance among the baselines. Compared with Homogeneous GNNs, AUC is at least 0.8\% increased and KS is at least 4.4\% increased, respectively. The benefits are derived from the incorporation of heterogeneous contents. However, the attribute information on links is not exploited, which is abundant in the BCN and conductive to the MSE default prediction. 
\end{itemize}

\subsection{Effects of Components (Q2)}

Next, we study the contributions of each component to further understand the traits of HIDAM. The evaluation results are summarized in Table~\ref{tab:ablation}.

\textbf{The effects of meta-paths}. We first analyze the benefits of meta-paths for characterizing default risks. Recall that our meta-paths depict semantic relationships in different views. The analysis is performed through removing the view-specific meta-paths, respectively denoted as HIDAM$_{\setminus F}$, HIDAM$_{\setminus I}$, HIDAM$_{\setminus E}$. For example, the variant HIDAM$_{\setminus F}$ excludes the meta-paths of fund view, including the $CtC$ path and $CtCtC$ path. Comparing with our full model, we can observe that AUC and KS are obviously decreased when removing any view-specific meta-paths, implying the effectiveness of composite relations captured by our proposed meta-paths. Among the results, the meta-paths of equity view have the most significant impact on predicting defaults in our dataset.

\textbf{The effects of link information}. Links involve various credit-related information, e.g. transactions between two companies. To verify the importance of link information in HIDAM, we remove all the link attributes and their associated operations (donated as HIDAM$_{\setminus L}$) for comparison. As shown in Table~\ref{tab:ablation}, we can observe that HIDAM$_{\setminus L}$ exhibits relatively unsatisfactory performance in prediction. The results indicate that link information has a valuable contribution for tackling the problem.

\textbf{The effects of attention mechanism}. By learning the importance of path instances and meta-paths, the integrated attention mechanism is to improve the representations of MSEs. To demonstrate the effectiveness of attention mechanism, we respectively remove the instance-level fusion (donated as  HIDAM$_{\setminus IF}$) and the semantic-level fusion (donated as  HIDAM$_{\setminus SF}$). The results shown in Table~\ref{tab:ablation} exhibit that HIDAM invariably outperforms its two variants, implying that two-level attention mechanism is capable to take advantage of the collected heterogeneous contents. Specifically, the instance-level fusion adaptively characterizes the diverse contributions of path instances to each MSE, rather than treating them equally. Then, the semantic-level fusion incorporates diverse semantics captured by meta-paths, which fully characterize default risks from different views. 

\renewcommand{\arraystretch}{1.2}
\begin{table}[b]
\caption{Results of component experiments, where HIDAM$_{\setminus *}$ denote a variant of HIDAM removing some component.}
\centering
\begin{tabular}{l|cc|cc}
\toprule
\multirow{2}{*}{\textbf{Variant}} & \multicolumn{2}{c|}{\textbf{AUC}} & \multicolumn{2}{c}{\textbf{KS}} \\
 \cline{2-5}
  & \textbf{Value} & \textbf{Effect}  & \textbf{Value} & \textbf{Effect} \\
\hline
HIDAM$_{\setminus F}$ & 0.7290 & $\downarrow$ 1.54\% & 0.3409 & $\downarrow$ 5.31\% \\
HIDAM$_{\setminus I}$ & 0.7276 & $\downarrow$ 1.73\% & 0.3382 & $\downarrow$ 6.06\% \\
HIDAM$_{\setminus E}$ & 0.7261 & $\downarrow$ 1.93\% & 0.3380 & $\downarrow$ 6.11\% \\
HIDAM$_{\setminus L}$ & 0.7280 & $\downarrow$ 1.67\% & 0.3398 & $ \downarrow$ 5.61\% \\
HIDAM$_{\setminus {IF}}$ & 0.7326 & $\downarrow$ 1.05\% & 0.3496 & $\downarrow$ 2.89\% \\
HIDAM$_{\setminus {SF}}$ & 0.7300 & $\downarrow$ 1.40\% & 0.3409 & $\downarrow$ 5.31\% \\
\midrule[1pt]
HIDAM & 0.7404 & - & 0.3600 & - \\
 \bottomrule
\end{tabular}
\label{tab:ablation}
\end{table}

In addition, almost all the variants work better than the baselines under the cases that some information or components are removed. These observations indicate that our adopted components for elaborately extracting and integrating heterogeneous information by meta-paths from different views can effectively capture MSE features for default prediction.

\subsection{Attention Analysis (Q3)}

\begin{figure}[t]
\centering
    \includegraphics[scale = 0.25]{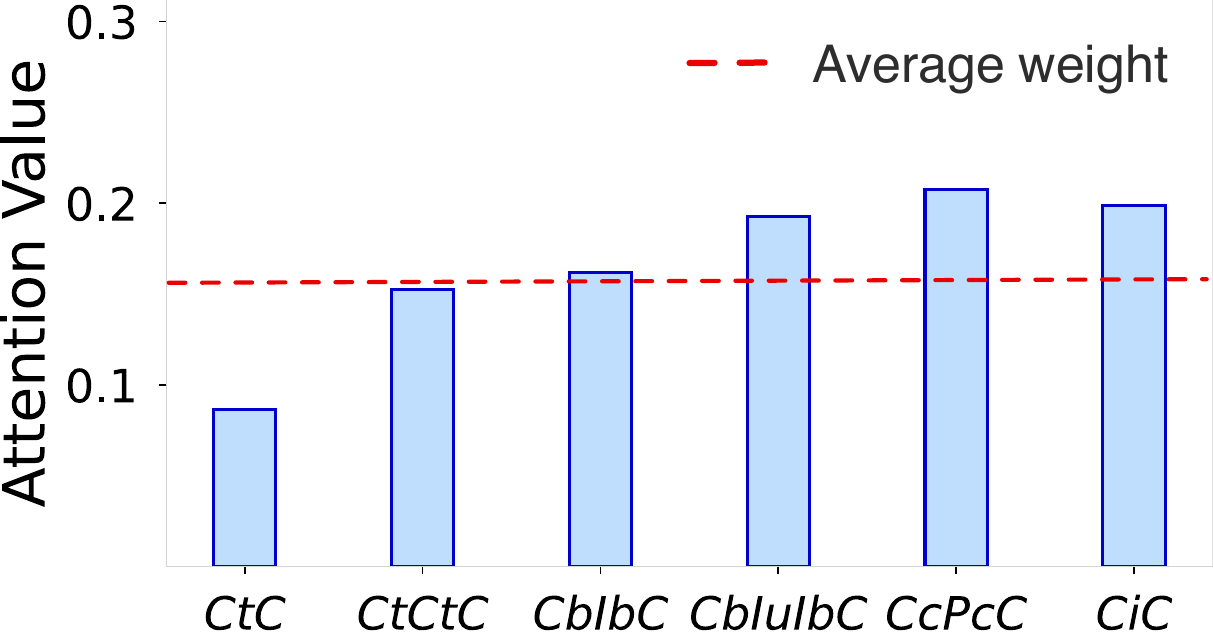}
\caption{The average attention values of different meta-paths.}
\label{fig5}
\end{figure}

We further analyze the attention mechanism in HIDAM. First, we report the average attentive weights of different meta-paths in the semantic-level fusion. As shown in Figure~\ref{fig5}, the distribution of the semantic attention values is relatively uniform. The observations reveal that the influence factors on MSE default in our dataset are complicated and all the meta-paths contribute to prediction. In addition, we can observe that the meta-paths from equity view (i.e. $CcPcC$ and $CiC$) contribute more weights than others, which conforms to our prior analyses.

\begin{figure}[t]
\begin{subfigure}{0.15\textwidth}
         \centering
         \includegraphics[scale = 0.14]{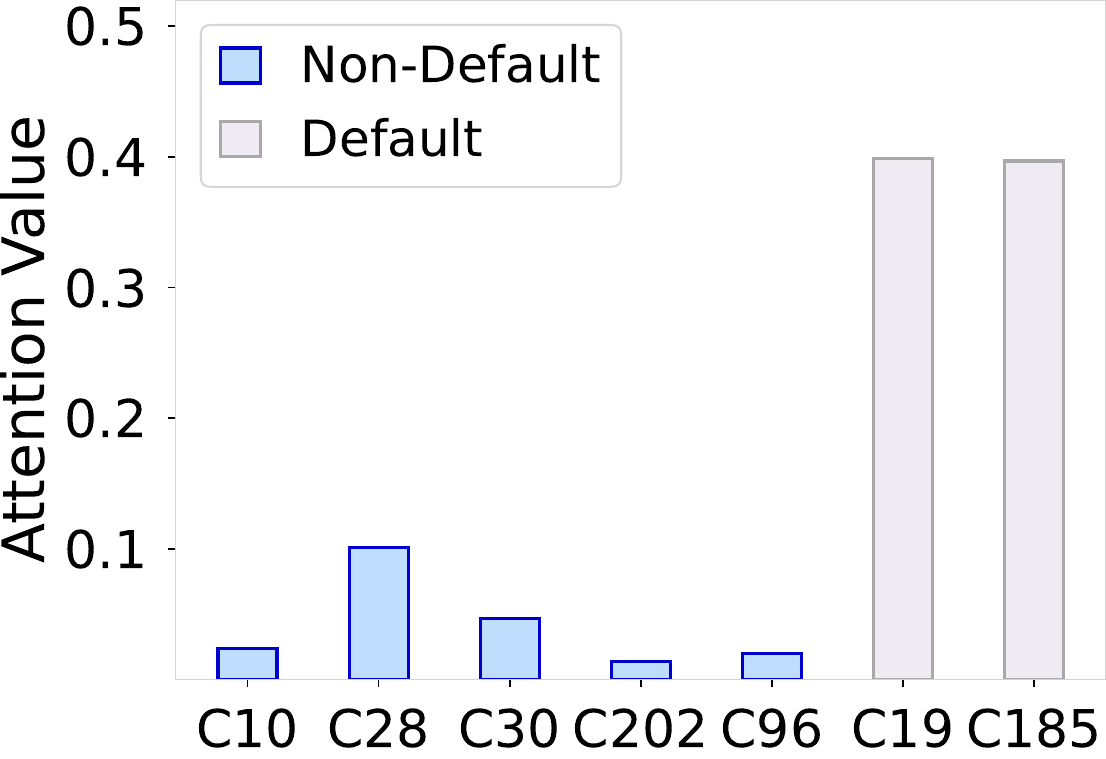}
         \caption{C136}
         \label{fig:6a}
\end{subfigure}\hspace*{\fill}
     \begin{subfigure}{0.15\textwidth}
         \centering
         \includegraphics[scale = 0.14]{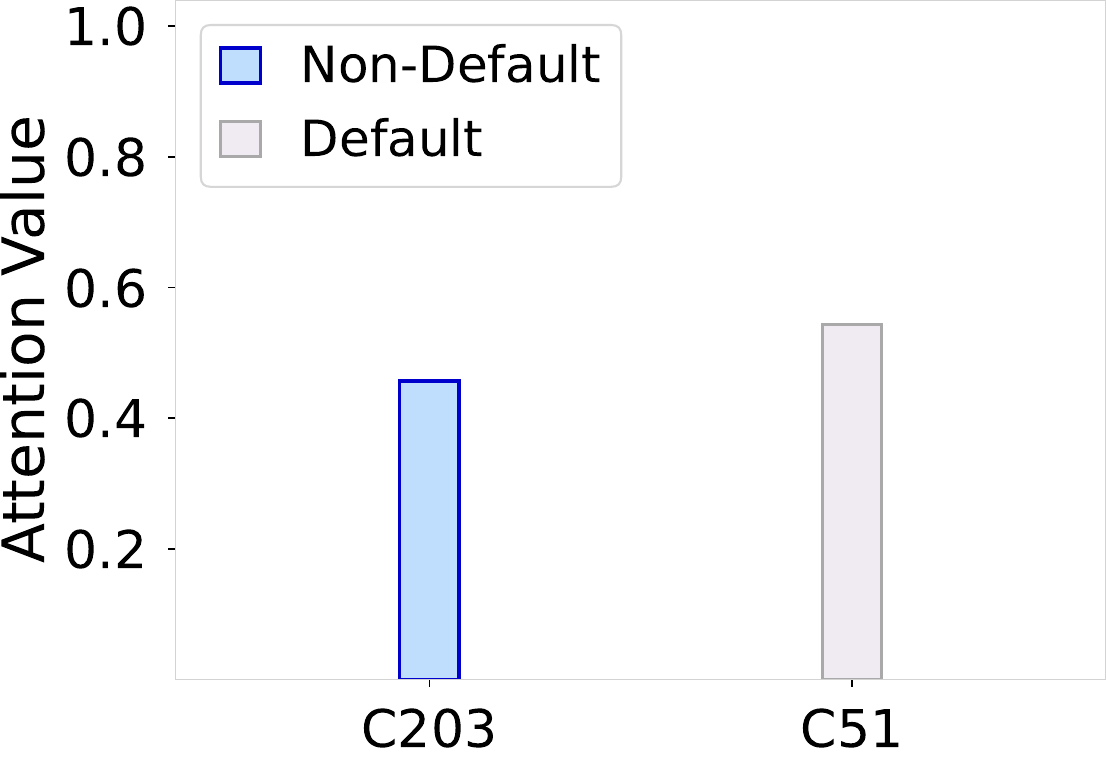}
         \caption{C81}
         \label{fig:6b}
     \end{subfigure}\hspace*{\fill}
    \begin{subfigure}{0.15\textwidth}
         \centering
         \includegraphics[scale = 0.14]{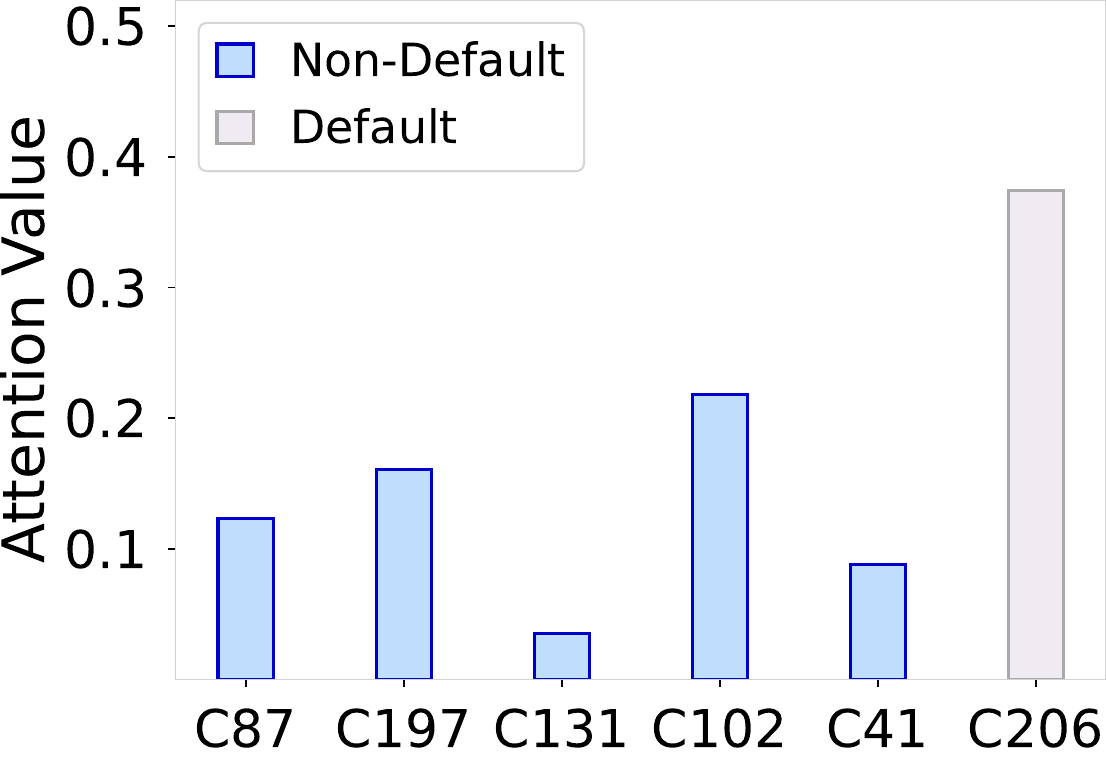}
         \caption{C175}
         \label{fig:6c}
     \end{subfigure}\hspace*{\fill}
\caption{An illustrative example of the instance-level fusion including three default MSEs.}
\label{fig6}
\end{figure}

Recall that the instance-level fusion can learn the importance of path instances. To reveal such capability, we give a microscopic analysis through a set of case studies including three default MSEs from the testing data, labeled as C136, C81, C175. Figure~\ref{fig6} shows their attention values in the instance-level fusion, where path instances are denoted as the corresponding meta-path based neighbors for simplification. We can observe that C19 and C185 get higher attention values for C136 in the $CbIbC$ path (see Figure~\ref{fig:6a}), C51 gets the highest attention values for C81 in the $CiC$ path (see Figure~\ref{fig:6b}), and C206 gets the highest attention values for C175 in the $CtC$ path (see Figure~\ref{fig:6c}). Note that all the above neighbors are labeled as default companies in our dataset, which offers an exogeneity related explanation for the three default MSEs. Thus, the observations support the effectiveness of HIDAM in modeling importance of path instances for the default prediction in the real-world networks.

\subsection{Parameter Sensitivity (Q4)}

\begin{figure}[t] 
\begin{minipage}{0.45\textwidth}
     \begin{subfigure}{0.48\textwidth}
         \centering
         \includegraphics[width=\textwidth,height = 2.5cm]{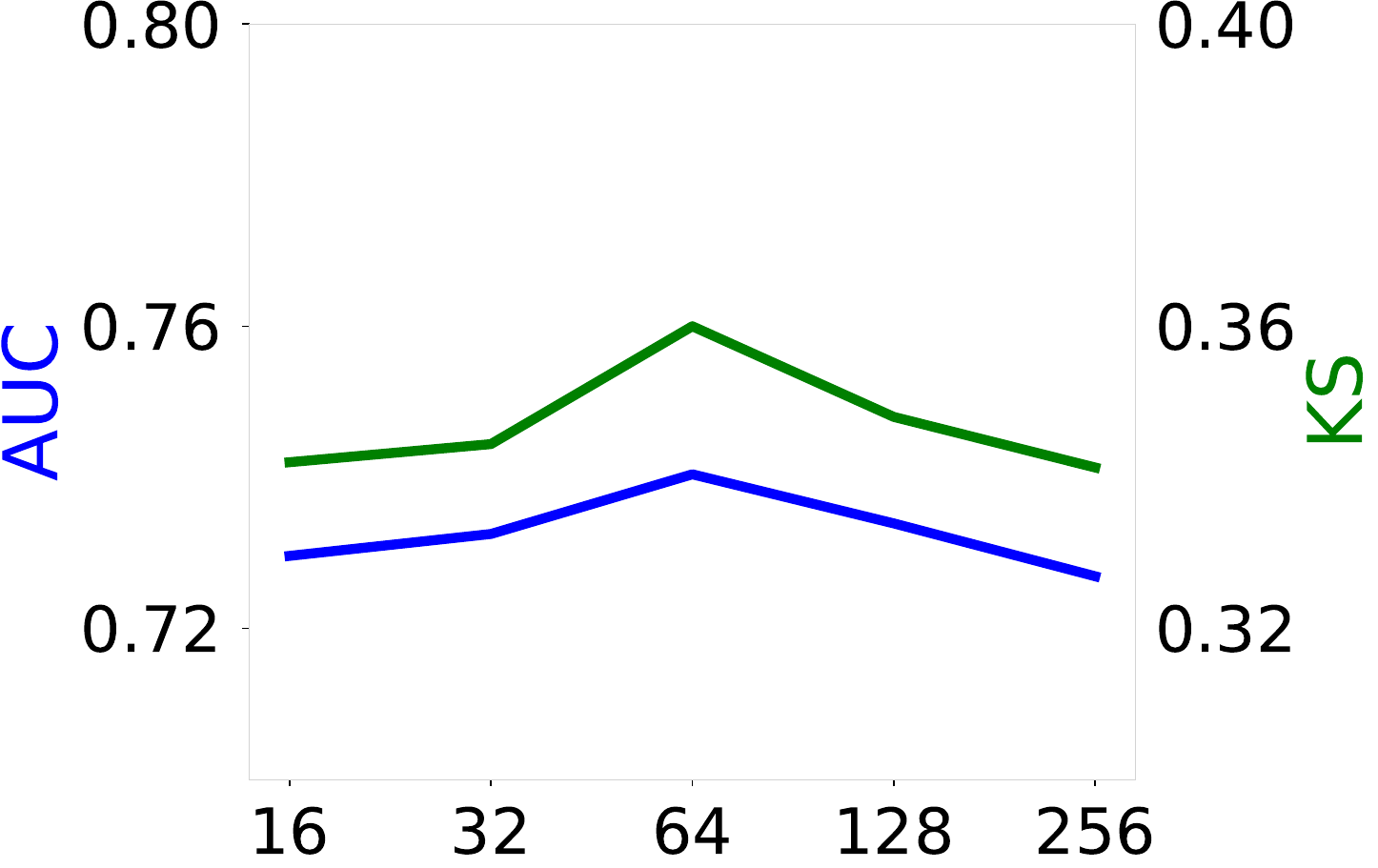}
         \caption{Dimension of the final embedding}
         \label{fig7a}
     \end{subfigure}\hspace*{\fill}
     \begin{subfigure}{0.48\textwidth}
         \centering
         \includegraphics[width=\textwidth,height = 2.5cm]{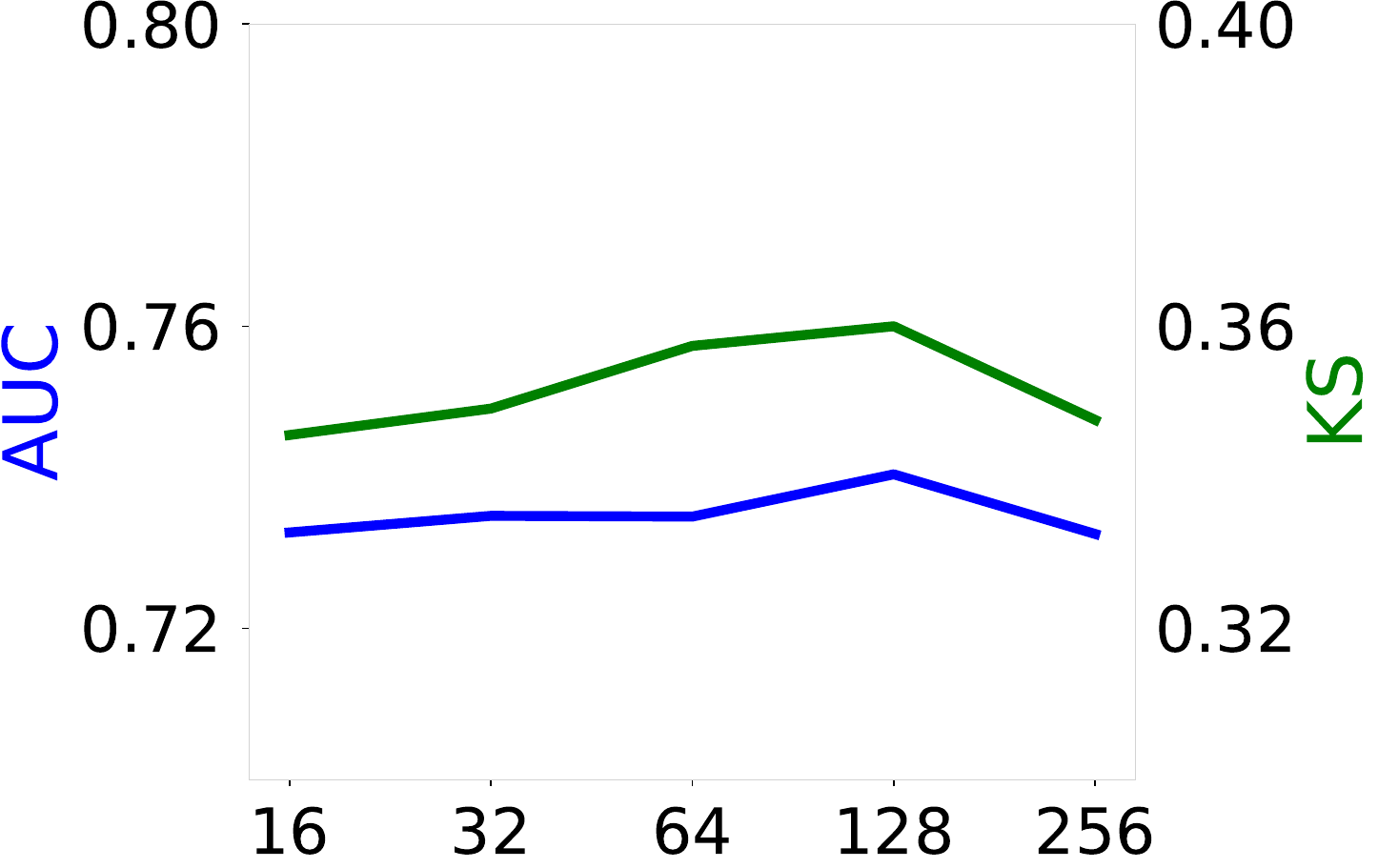}
         \caption{Dimension of the semantic-level attention vector}
         \label{fig7b}
    \end{subfigure}\hspace*{\fill}
\end{minipage}
\caption{Performance of our model on default prediction under different settings.}
\label{fig7}
\end{figure}

To study the impact of parameters, we examine the parameter sensitivity of HIDAM through tuning a single parameter while keeping others unchanged. Figure~\ref{fig7} exhibits the experimental results of different parameter settings. First, we study how the dimension of the final embedding influences prediction. As shown in Figure~\ref{fig7a}, when increasing the embedding dimension, both AUC and KS rise and then drop. The results indicate that an appropriate dimension value can encode sufficient information but a larger value may result in overfitting and redundancy. Then, we report the results of adjusting the dimensions of the semantic-level attention vector in Figure~\ref{fig7b}. We can observe that the model presents a similar but smoother performance change trend. The AUC and KS values fluctuate within a small range, which indicates these dimensions are enough to capture semantic preferences. By comparison, our model is more sensitive to the embedding dimension.

\section{Conclusion}

In this paper, we study the problem of graph-based financial default analysis for MSEs in commercial banks. By elaborately observing the relational data involved in financial activities and default cases, we propose a novel HIDAM model to tackle the problem. Specifically, we attempt to model the scenario of business banking service as an attributed heterogeneous information network. For better representation of MSEs, we aggregate interactive information through meta-paths, and then exploit both node feature contents and intermediate messages along the meta-path. Furthermore, we design a two-level attention mechanism to learn the importance of path instances and meta-paths simultaneously. Experiments on real-world banking datasets verify that HIDAM is effective to financial default prediction for MSEs. As future work, we will continue to incorporate more credit-related data and further demonstrate the effectiveness of HINs on the business banking service and finance risk 
problems, e.g. guaranteed loan, debt note, supply chain finance. Another attempt is to generate more comprehensible default reasons by further exploiting the interpretability of HINs.

\bibliographystyle{ACM-Reference-Format}
\bibliography{sample-base}

\end{document}